\documentclass[runningheads]{llncs}
\usepackage[T1]{fontenc}
\usepackage{graphicx}
\usepackage{hyperref}
\usepackage{color}

\usepackage{xspace}
\newcommand{\ourtool}{TRACE\xspace}
\graphicspath{{figures/}}
\usepackage{todonotes}
\usepackage{subcaption}
\usepackage{xcolor}
\usepackage[misc]{ifsym}

\begin{document}
\title{Pattern or Artifact? Interactively Exploring Embedding Quality with TRACE}
\toctitle{Pattern or Artifact? Interactively Exploring Embedding Quality with TRACE}
\titlerunning{Interactively Exploring Embedding Quality with TRACE}
\author{Edith Heiter\inst{1}\Letter\orcidID{0000-0003-0279-2139} \and
Liesbet Martens\inst{1,2}\orcidID{0000-0001-9180-7456} \and
Ruth Seurinck\inst{1,2}\orcidID{0000-0002-6636-7572} \and
Martin Guilliams\inst{1,2}\orcidID{0000-0003-3525-7570} \and
Tijl De Bie\inst{1}\orcidID{0000-0002-2692-7504} \and
Yvan Saeys\inst{1,2}\orcidID{0000-0002-0415-1506} \and
Jefrey Lijffijt\inst{1}\orcidID{0000-0002-2930-5057}}
\tocauthor{Edith Heiter, Liesbet Martens, Ruth Seurinck, Martin Guilliams, Tijl De Bie, Yvan Saeys, Jefrey Lijffijt}
\authorrunning{E. Heiter et al.}
\institute{Ghent University, Ghent, Belgium\and
	VIB-UGent Center for Inflammation Research, Ghent, Belgium\\
	\email{edith.heiter@ugent.be}}

\maketitle              %
\begin{abstract}
This paper presents \ourtool, a tool to analyze the quality of 2D embeddings generated through dimensionality reduction techniques. Dimensionality reduction methods often prioritize preserving either local neighborhoods or global distances, but insights from visual structures can be misleading if the objective has not been achieved uniformly. \ourtool addresses this challenge by providing a scalable and extensible pipeline for computing both local and global quality measures. The interactive browser-based interface allows users to explore various embeddings while visually assessing the pointwise embedding quality. The interface also facilitates in-depth analysis by highlighting high-dimensional nearest neighbors for any group of points and displaying high-dimensional distances between points. \ourtool enables analysts to make informed decisions regarding the most suitable dimensionality reduction method for their specific use case, by showing the degree and location where structure is preserved in the reduced space. 


\end{abstract}
\begin{figure}
\includegraphics[width=\textwidth]{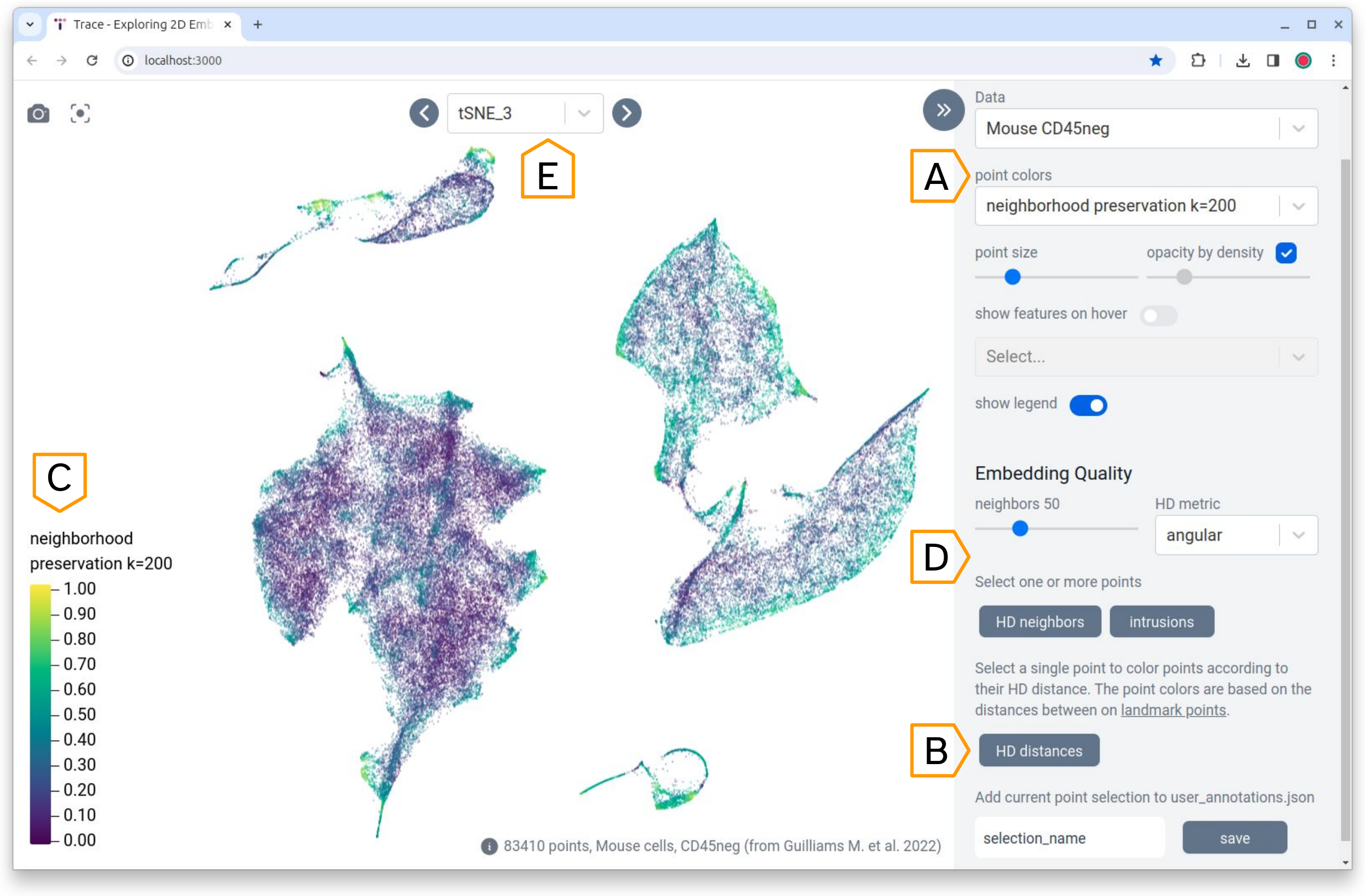}
\caption{The \ourtool user interface showing an embedding where points are colored according to neighborhood preservation.}
\label{fig:interface}
\end{figure}

\section{Introduction and Related Work}
Visualizing high-dimensional data in 2D is an important step of data understanding as it helps to grasp the structure of the dataset. Dimensionality reduction methods that are used to compute such embeddings inevitably introduce distortions and cannot represent all structure in the data faithfully. While these techniques are benchmarked based on average embedding quality, the results are rarely visualized on the embedding itself. We argue that  visualizing the embedding quality for each point is a helpful step in dimensionality reduction to (a) understand the global structure of the dataset which might not be fully retained in the embedding, (b) identify where local neighborhood distortions occur, and (c) efficiently compare embeddings to observe where they differ in terms of quality.
Existing Python libraries such as pyDRMEtrics \cite{zhang2021pydrmetrics} or ZADU \cite{jeon23vis} offer functionality to measure embedding quality but lack interactive visualizations. Tools that are specifically designed for quality visualization struggle to scale to large datasets due to their reliance on pairwise distance measures  \cite{lespinats2011checkviz,heulot2013proxilens,chatzimparmpas2020t}.

\paragraph{\textbf{Contributions}} To address these limitations, we developed a Two-dimensional Representation Analysis and Comparison Engine (\ourtool) with a flexible and \textbf{extensible} backend in Python. The implementation includes \textbf{scalable} quality measures as well as an \textbf{interactive} user interface. The source code is available on \href{https://github.com/aida-ugent/TRACE}{github.com/aida-ugent/TRACE} and a demo video on \href{https://youtu.be/mtyFzXt51Jw}{youtu.be/mtyFzXt51Jw}.

\section{Three Use Cases for \ourtool}
\paragraph{1. To understand the global structure preservation,} points can be colored based on the distance rank correlation or random triplet accuracy \cite{wang2021understanding} (Figure \ref{fig:interface} {\color{orange}A}). This allows the analyst to identify where the relative placement of points is faithful to the original high-dimensional data.
Moreover, by changing point colors to represent (approximate) high-dimensional distances to any selected point ({\color{orange}B}), the actual global arrangement of substructures can be explored in detail. This is particularly useful for points with low global quality scores, as it reveals the points that are closer or more distant than their relative positions in the embedding suggest. 

\paragraph{2. Analyzing local distortions} by visualizing the local neighborhood preservation \cite{lee2009quality} for the precomputed neighborhood sizes ({\color{orange}C}). This helps determine how well the nearest neighbors in the embedding align with the nearest neighbors in the high-dimensional space. Neighborhood preservation scores often vary across the embedding, with interesting local differences. Depending on the local quality score, one could either be reassured or discouraged from relying on visual patterns such as small clusters or linear shapes in subsequent analyses. To understand why a set of points has low or high neighborhood preservation scores, the analyst can visualize (the union of) the high-dimensional neighbors ({\color{orange}D}). 

\paragraph{3. Comparing embeddings} that preserve different aspects of the data, while being aware about their quality trade-offs, contributes to a deeper understanding of the data. Highlighting points with highly variable placement across embeddings can be achieved using point stability scores \cite{sun2023dynamic}. 
Tracing these points through different embeddings ({\color{orange}E}), can reveal valuable insights into their subtle differences.

\begin{figure}[t]
	\includegraphics[width=\textwidth]{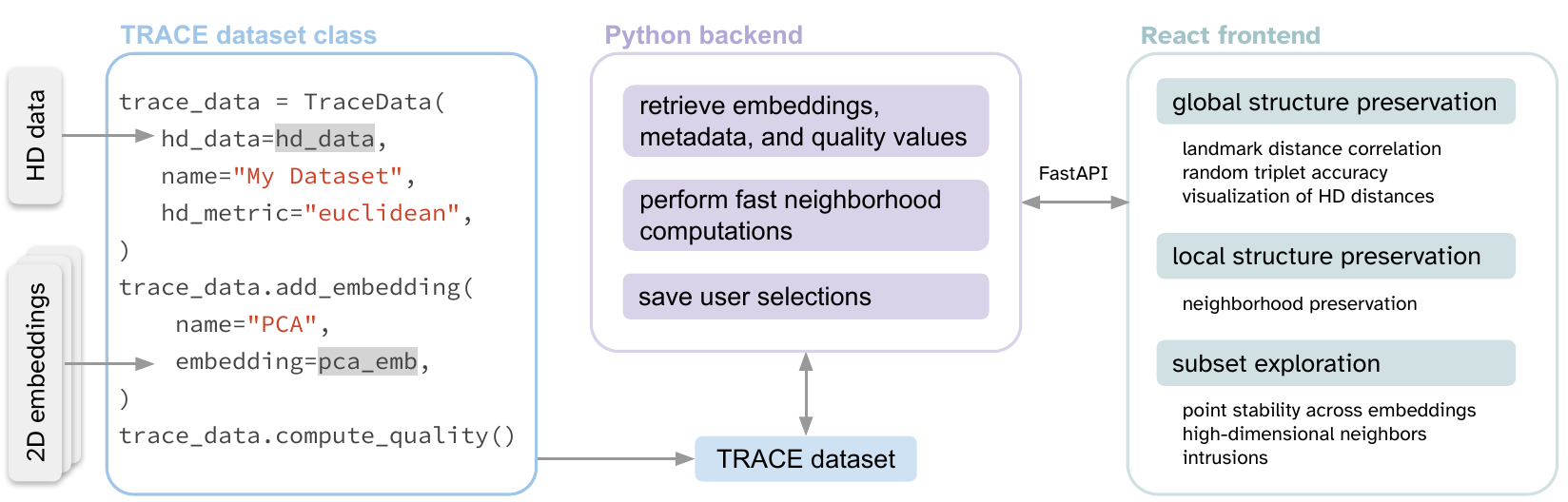}
	\caption{Implementation overview of \ourtool where embeddings and quality measures are precomputed before loading them with the backend.}
	\label{fig:pipeline}
\end{figure}

\section{Design Choices and Implementation}
\ourtool can to be used with larger datasets as we separated the computation of 2D embeddings and quality measures from the interactive visualization. This allows for efficient processing: if the embeddings are available, all quality measures will be precomputed with sensible default values using just one line of code (Figure \ref{fig:pipeline}). The dataset class is extensible, enabling users to implement additional quality measures that can use the precomputed high-dimensional neighbors.

The user interface (Figure \ref{fig:interface}) focuses on one visualization at a time while allowing for intuitive switching between different embeddings and point colors. This interactive design is built upon regl-scatterplot \cite{lekschas2023regl}, which provides (lasso) selection, panning, zooming, and smooth transitions between embeddings as standard features. To optimize performance, only the data necessary to visualize the selected embedding is fetched from the backend.

\begin{figure}[t]
	\includegraphics[width=\textwidth]{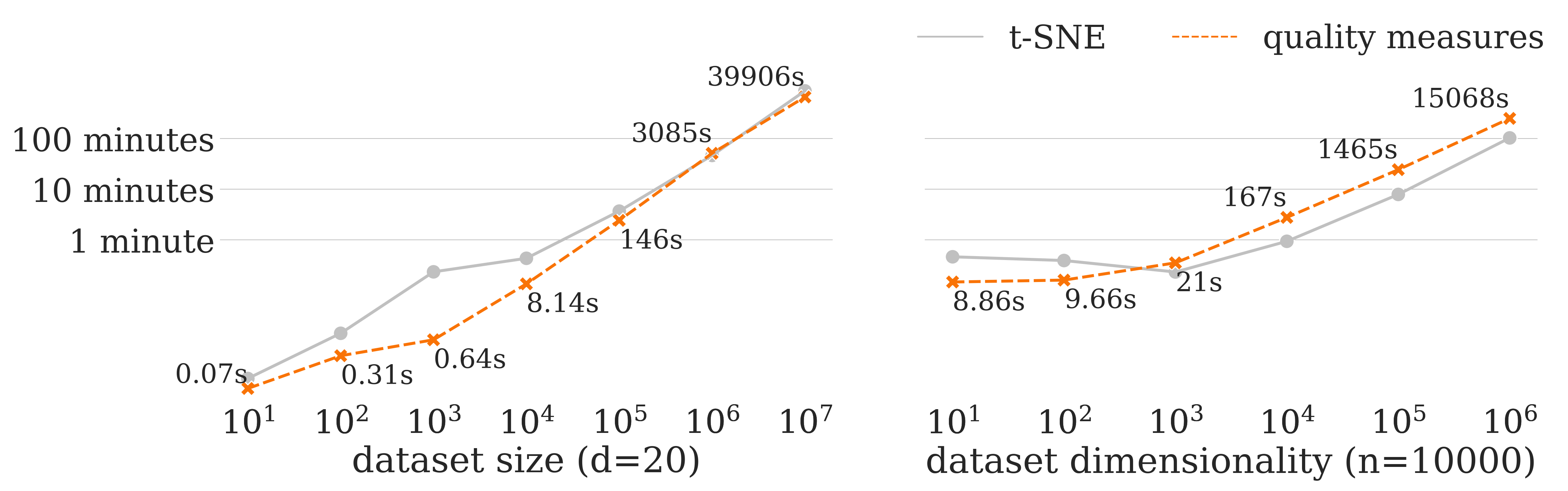}
	\caption{The time to compute all quality measures for five embeddings is comparable to running t-SNE (perplexity 30, 800 iterations). We used an Intel Xeon Gold 6136 CPU, using a maximum of 12 cores.}
	\label{fig:benchmark}
\end{figure}

To improve scalability, we incorporated design choices such as approximate nearest neighbors and parallelizing quality measure computation with Numba. As depicted in Figure \ref{fig:benchmark}, quality measure calculation is as efficient as generating a low-dimensional embedding, as demonstrated by benchmarking on datasets with varying sizes and dimensions.

\begin{credits}
\subsubsection{\ackname} 
This research was funded by the BOF of Ghent University (BOF20/IBF/117), the Flemish Government (AI Research Program), and the FWO (project no. G0F9816N, 3G042220, G073924N, 11J2322N).
\subsubsection{\discintname} The authors have no competing interests to declare that are relevant to the content of this article. 
\end{credits}
\bibliographystyle{splncs04}

\end{document}